\documentclass[aps,jcp,amsmath,amssymb,reprint]{revtex4-2}

\usepackage{chemformula} % Formula subscripts using \ch{}
\usepackage[T1]{fontenc} % Use modern font encodings
\usepackage{multirow}
\usepackage{dcolumn}
\usepackage{tabularx}
\usepackage{float}
\usepackage{xr}
\usepackage{hyperref}
\usepackage[utf8]{inputenc}
\usepackage{mciteplus}
\usepackage{subfig}
\usepackage[utf8]{inputenc}
\setcitestyle{super}
\usepackage[cdot,mediumqspace,amssymb]{SIunits}

\usepackage{comment} % to comment out your intro so you can work on it.

\usepackage{textcomp} % to get a tilde
\usepackage{threeparttable}

\begin{document}
	\preprint{AIP/123-QED}
	
	\title{Effective core potentials as a pathway to self-interaction error correction: a proof-of-concept study on one-electron systems}
	
	\author{Dale R. Lonsdale}
	\email{dlonsdale@outlook.com.au}
	\author{Lars Goerigk}%
	\email{lars.goerigk@unimelb.edu.au}
	%\phone{+61 3 834 46784}
	\affiliation{School of Chemistry, The University of Melbourne, Victoria 3010, Australia}

\begin{abstract}

In all applications of Density Functional Theory there is always a degree of one-electron self-interaction error (SIE).
Here, we propose a simple self-interaction correction by applying an effective core potential (ECP) that replaces no electrons: we dub this the self-interaction potential (SIP).
ECPs are already implemented in all major quantum chemistry codes and so there is minimal effort required by developers and users to access our correction.
The goal of SIPs is to reduce the overly-severe SIE --- commonly manifesting as the unphysical delocalization of an electron over two or more nuclei, or even over an entire chemical system. 
We propose two first generation SIPs (optimized SIPs and subtraction SIPs) that can reduce the SIE in various one-electron test systems and a hydrogen transfer reaction. Our tests show improvements for systems that suffer from predominantly functional- or density-driven errors, pointing at the potential robustness of the approach. 
Herein, the viability of SIPs is demonstrated in a proof-of-concept study and several avenues for improvement are identified that will help with the construction of future generations of SIPs.

\end{abstract}
\maketitle

% \input{./sections/introduction.tex}
\section{Introduction}
\label{sec:intro}

Kohn-Sham (KS) Density Functional Theory\cite{KSDFT_orig,Hohenberg1964} (DFT) is frequently used for molecular chemistry and solid state physics as a tool to predict and explain chemical phenomena.  
Successful calculations to the average user means obtaining a result that is "close enough" to the true answer and, just as critically, in a meaningful time frame. 
Such desires are often stymied by the shortcomings of DFT, in particular, the self-interaction error (SIE).\cite{perdew1981self, Lonsdale2020, Lonsdale2023}
Specifically, it is the cause of incorrect predictions of bond dissociation,\cite{Chermette2001, Grfenstein2004} charge transfer,\cite{Ruiz1996, Grfenstein2004} polarizability in molecular chains,\cite{Kmmel2004, Ruzsinszky2008} band gaps,\cite{MoriSnchez2008, Singh2017} excitation energies,\cite{Goerigk2008, Goerigk2009, Goerigk_2010} and barrier heights.\cite{H3_barrier_underestimate, Andersson2004}
SIE can be so large in magnitude that the DFT calculation is qualitatively incorrect, failing to provide any meaningful insight.
A partitioning of the SIE into many-electron and one-electron components has been done by Mori-S{\'{a}}nchez, Cohen, and Yang.\cite{2006_many_elecsie} 
While both components can  play a role in the above mentioned deficiencies, our present work focuses only on the latter.

Self-interaction corrections (SICs) were utilized as early as 1930 by the Hartree-Fock (HF) method, in which the Coulomb self-interactions found in the Hartree method\cite{Hartree1928} are implicitly corrected for by accounting for the antisymmetry of the wave function that introduces (equally-sized) self-exchange terms.\cite{Fock1930}
In the development of DFT, the first SIC comes from Fermi and Amaldi (1934) whereby they developed a rudimentary correction to the early Thomas-Fermi model.\cite{FERMI_AMALDI_SI_correction} 
For DFT the most-used one-electron SIE correction is likely that developed by Perdew and Zunger in 1981.\cite{perdew1981self} 
The Perdew Zunger (PZ) correction removes the SIE orbital by orbital, but in its application the original PZ-SIC tends to over-correct the total energy and in such cases leads to results that are worse than the uncorrected.\cite{Goedecker1997, SIC_BAD_thermo_2004, Vydrov2005, Akter2021}
Additionally, the PZ-SIC is not invariant with respect to unitary transformation:

\begin{equation}\label{eq:PZ-SIC}
E^{PZ} = E^{total} - \sum^n_i(J_{ii} + E_{xc}[\rho_i])\ ,
\end{equation}

where $E^{PZ}$ is the Perdew Zunger-corrected energy, $E^{total}$ is the total energy calculated by a KS density functional approximation (DFA), $J_{ii}$ is the self-Coulomb interaction of the $i^{th}$ orbital, $n$ is the total number of occupied orbitals, and $E_{xc}[\rho_i]$ is the energy from the exchange correlation functional corresponding to the electron density of the $i^{th}$ orbital. 
The way in which one defines the orbitals will change the magnitude of the correction --- thus it becomes an orbital-dependent energy.
In contrast, unitary transformations of the orbitals from a true Kohn-Sham density functional do not change the total energy.
As a result, PZ-SIC functionals are not as simple to apply as KS-DFT --- the increased complexity of an orbital-dependent optimization leads to increased computational overheads and difficulties when conducting orbital-based analyses.

Thus, many sought to improve on one or more of the drawbacks of the original PZ-SIC: using an optimized effective potential (OEP)\cite{Sharp1953, Talman1976} known as OEP-SIC\cite{Krzdrfer2008} for ground and TD-OEP-SIC\cite{Chu2004} for excited states; using the Krieger-Li-Iafrate (KLI)\cite{Li1993} OEP approximation for ground (KLI-OEP-SIC),\cite{Tong1997} and excited states (TD-KLI-OEP-SIC);\cite{Ullrich1995, Petersilka2000} scaling down the PZ correction;\cite{Vydrov2006_scaled_down_SIC, Vydrov2006_simplified_scaled_SIC} utilizing either Fermi-localized orbitals (FLO-SIC)\cite{FERMI_SIC_2014, Fermi_derivatives_Pederson2015, FLOSIC_pastLDA} or complex orbitals;\cite{Hannes_First_cmplxSIC, complex_SIC_cite_first_Lehtola2016, complex_SIC_cite_second_Lehtola2016} incorporation of pseudopotentials;\cite{Vogel1996, filippetti2003, ASIC2007, Filippetti2011} or creation of Koopmans-compliant functionals.\cite{Dabo2010,Borghi2014}
Moving away from the traditional PZ-correction include the Hubbard U correction (DFT+U) for transition metals systems\cite{1991dft+u_vladimir, Kulik2006_dftU,KirchnerHall2021}, and density corrected DFT (DC-DFT).\cite{Lee2010, Sim2014, Sim2018} The latter is based on the insights that the many-electron SIE can be separated into a functional and density-driven component, with DC-DFT addressing the latter. For the one-electron component of the density-driven error, see our definition in Ref. \citenum{Lonsdale2020}.

The above mentioned PZ-SIC extensions are not yet commonplace. Instead, the \textit{de facto} solution to the SIE is to use DFAs with an admixture of exact exchange, such as global hybrids,\cite{Becke1993} global double hybrids\cite{B2PLYP} or functionals with range separation.\cite{Savin1995, Leininger1997} Note that the latter was shown to be anything but a guaranteed solution to the one-electron SIE.\cite{Lonsdale2020} 

In the solid-state physics community, using variations of the DFT+U correction on transition metals to correct the SIE is standard due to the computational expense of hybrid level functionals in such applications.\cite{Dudarev1998, Kulik2006_dftU, Trimarchi2018, Varignon2019, KirchnerHall2021}
We observe a missing gap --- an easy-to-use SIC for the lower rungs of Jacob's Ladder in molecular-based calculations.

In the spirit of an efficient SIC, we have developed the beginnings of such a correction based on re-purposing effective core potentials (ECPs). 
ECPs were originally designed to replace core orbitals/electrons on heavier elements by an effective potential and thus reduce computational effort.\cite{ECP1,ECP2,ECP3,ECP4,ECP5}
Additionally, one can fit scalar relativistic effects into the ECP.\cite{ECP4,ECP5,Dolg2012}
Contrary to the intended use of ECPs, our desire is to use these potentials on "top" of the system, i.e. we do not remove the explicit calculation of any core orbitals.
We are not the only ones to appropriate ECPs in such a way, atom-centered potentials (ACPs) from DiLabio and coworkers are most similar to our own. 
Quantum-capping ACPs (2002), in which an ECP that replaces a single valence electron is combined with a shielding potential to model atoms at the partition of a quantum mechanical/molecular mechanical boundary, are the earliest.\cite{DiLabio2002}

Mimicking London dispersion effects was attempted by re-purposing ECPs, such as in approaches dubbed ``dispersion corrected atom centered potentials" (DCACPs),\cite{vonLilienfeld2004} ``local atomic potential'',\cite{Sun2008} DCACP2,\cite{Karalti2014} and different variants of ``dispersion corrected potentials'' (DCPs).\cite{DiLabio2008first,Johnson2009_second,b3lyp-dcp} Note that the inability of the latter to capture intramolecular dispersion in conformers and thermochemistry was demonstrated in 2014.\cite{Goerigk2014}

Additional uses for refitted ECPs include: force-corrected atom centered potentials for geometries and harmonic frequency calculations,\cite{Lilienfeld2013_FCACP} torsion-corrected atom centered potentials,\cite{Tahchieva2018} whereby a pseudopotential is optimized to yield better rotational barriers, potentials to improve modeling water clusters,\cite{Holmes2017} and ACPs to be used as corrections in small-basis set calculations for DFT and HF.\cite{Prasad2018, Prasad2022_march, Prasad2022_april}

ECPs, by the name of ``pseudopotentials'', have been used in SIC calculations before. First by Zunger in 1980\cite{zunger1980_SIC_pseudo-corrected} and later again by several others in various forms through the 1990s to the current day.\cite{Rieger1995, Vogel1996, Baumeier2006, filippetti2003, ASIC2007, Filippetti2011}
Our usage of ECPs differs to these pseudopotential-SIC methods in three ways: 1) we do not use the PZ-SIC formalism in any way, 2) we intend our ECP to be used on molecular calculations as opposed to the solid state, and 3) we do not replace any electrons in the calculation. We refer to our ACP-style correction as a ``self-interaction potential" with the acronym ``SIP". 

ECPs are already implemented in all major quantum chemistry codes and thus little additional effort is expended to apply a SIP --- a major advantage in terms of applicability over many other SIC methodologies.
With either no or minimal modification to pre-existing code, users have the freedom to choose whatever software they desire.
Computational cost of applying an ECP is similarly minimal.
We hope then to satisfy the construction of a low-cost SIC that can be used to correct lower-rungs of Jacob's ladder, but potentially also improve accuracy of even sophisticated functionals.

\begin{figure*}
\begin{center}
\includegraphics[width=0.75\linewidth]{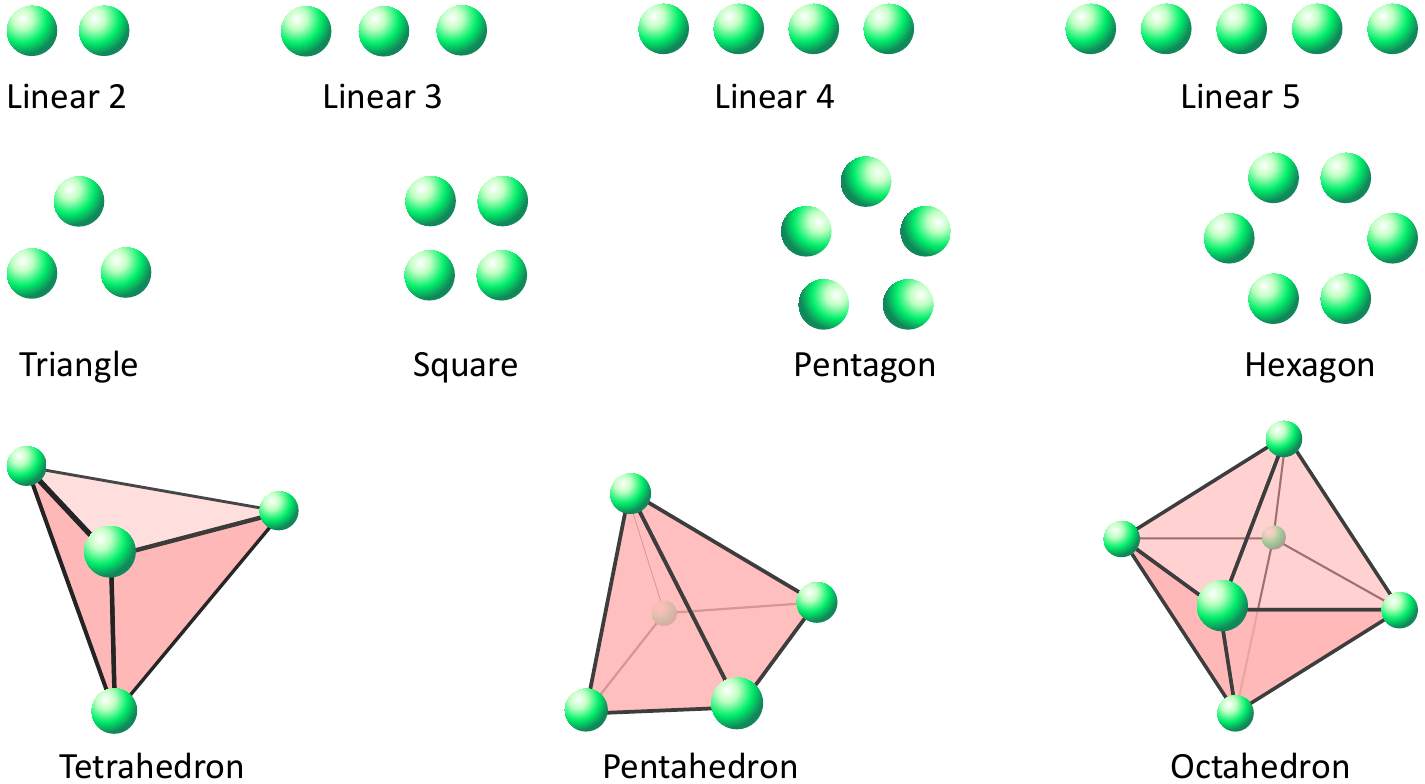}
\end{center}
\caption{Model systems for our one-electron calculations on various geometries; figure reproduced from Ref. \citenum{Lonsdale2023} (Creative Commons CC BY license).}
\label{fig:geometries}
\end{figure*}

Starting from the results of our previous two papers,\cite{Lonsdale2020,Lonsdale2023} in which we studied the SIE using mostly one-electron model systems, we discuss the findings separately. 
Our first paper was mostly a benchmark study of 74 density functionals on one-electron mono- and dinuclear systems, in fact the largest one-electron SIE benchmark study conducted.\cite{Lonsdale2020}
We showed how one-electron calculations provided similar analysis as the usually conducted fractional electron number studies.  
Some ``SIE-free'' functionals where shown to actually be examples of error compensation, which broke down when one-electron mononuclear systems other than the hydrogen atom were considered. 
Also established was how the one-electron SIE was linear with increasing nuclear charge $Z$, that the SIE had a basis-set dependence, and that van-der-Waals functionals of the VV10 type\cite{Vydrov2010} suffer from ``self-dispersion''.
Contrary to the belief of some, range-separation was shown not to be a `silver bullet' to the one-electron SIE --- despite the simplicity of our model systems, a non-negligible SIE remained in range-separated functionals. 
Finally, we also pointed out the case of positive SIE in dissociation curves and as a potential cause of incorrectly calculated bond lengths.

The second work extended upon the first by inclusion of one-electron systems with different one-, two- and three-dimensional geometries (e.g. a triangle, hexagon, and octahedron among others) and an analysis into the SIE of orbital occupation higher than the 1s orbital.\cite{Lonsdale2023} 
Geometries from this work are shown in Fig. \ref{fig:geometries}.
We showed that larger geometries (one electron delocalized over more nuclei) have larger SIEs, as does occupation in the 2s and 2p valence orbitals. 
Additionally, asymmetric fractional electron occupation was linked to a higher density error as opposed to symmetric fractional electron occupation, i.e. if all nuclei have the same amount of fractional electron the SIE is functional-error dominated, instead.

In the following section we briefly describe the theoretical background and the idea behind the construction of SIP libraries, for which we partially relied upon the geometries introduced in our second article. 
The idea behind a SIP library is that the DFT user gets an estimate for the one-electron SIE in their system and then chooses the right SIP that should correct this SIE, which is then used in the actual self-consistent field (SCF) calculation. 
The performance of these libraries is then evaluated on various benchmark sets --- mostly our established one-electron model systems, but also to a simple hydrogen transfer reaction. 

Note that this work serves as a proof-of-concept study on the use of SIPs to correct the one-electron self-interaction error in hydrogen-based test systems. Based on our findings and conclusions, further developments will be enabled to more comprehensively address in the one-electron SIE.

\section{Theoretical background}\label{subsec:theoretical_background}

Self-interaction comes from the Coulomb interaction:

\begin{equation}
J[\rho] = \frac{1}{2} \iint \frac{\rho(\vec{r}_1)\rho(\vec{r}_2)}{\vec{r}_{12}} \ d\vec{r}_1d\vec{r}_2 \ \ ,
\label{eq:coulomb_energy}
\end{equation}

where $\rho$ is the electron density, $\vec{r}$ represents the spatial coordinates of an electron, and $\vec{r}_{12}$ is the distance between two electrons. 
While not obvious, Equation \ref{eq:coulomb_energy} contains a fictitious self-interaction energy in which an electron `feels' a repulsion to its own density.
Experimentally this repulsion does not exist and in exact KS-DFT is exactly canceled in the true density functional:

\begin{equation}\label{eq:KS-DFT}
E[\rho] = T_s [\rho] + \int d\textbf{r}\ \nu_{ext}(\textbf{r})\rho (\textbf{r}) + J[\rho] + E_{XC}[\rho]\ \,
\end{equation}

where $T_s$ is the kinetic energy of the non-interacting electrons, $\nu_{ext}$ is the external potential energy, $E_J$ is the Coulomb energy, and $E_{XC}$ is the exchange-correlation energy. The exact functional expression of the latter is unknown and needs to be approximated, but it formally describes all exchange and correlation effects, corrects for the usage of non-interacting electrons in the treatment of the kinetic energy, and cancels the self interaction stemming from $E_J$. Due to using an approximate $E_{XC}$ expression, any remnants of the resulting imperfect cancellation between self-interaction in $E_J$ and  $E_{XC}$ is referred to as the SIE.

In one-electron systems the SIE is an easy value to quantify, which we do later.
Many-electron systems contain a one-electron SIE and a many-electron SIE. 
The orbital-based Perdew-Zunger correction as per Eq. \ref{eq:PZ-SIC} attempts to quantify the one-electron SIE in many-electron systems.
However, writing down an equation to quantify the many-electron SIE is non-trivial, and in this study we focus predominantly on one-electron systems.

Our correction is based on the standard definition of an ECP --- as a linear combination of Gaussian-type functions:

\begin{equation}\label{eq:SIP_equation}
U_l[r] = r^{-2} \, \sum_{i=1} c_{li} r^{n_{li}} e^{-\zeta_{li}r^2} \  , 
\end{equation}

where $r$ is the distance from the nucleus, $l$ is the angular momentum quantum number, $n_{li}$ is the power of the radial pre-factor, which is set to 2 in our case, $c_{li}$ is the contraction coefficient, and $\zeta_{li}$ is the Gaussian exponent.
We then optimize or choose the Gaussian exponents and contraction coefficients so that the ECP removes a portion of the one-electron SIE.

\section{Technical details}
\label{subsec:technical_details}

\subsection{Self-interaction potential libraries}
\label{subsec:SIP_libraries}

Currently SIPs are a library --- for a particular SIE value one extracts the appropriate SIP from a catalog of exponent-coefficient pairs.
How to determine the ``appropriate'' SIP depends on which type of library it comes from.
We have two approaches in constructing our SIP libraries: optimized SIPs (O-SIPs) and subtraction SIPs (S-SIPs). 

O-SIPs have their contraction coefficients and Gaussian exponents optimized via a least squares regression to minimize the SIE from a chosen set of reference systems. 
Currently this library type is made from a single reference system at a range of internuclear separations, e.g. a new SIP for each point on a dissociation curve of H$_2^+$.
Choosing an appropriate SIP to remove the SIE from any given test system is done by matching an estimation of the SIE of the test system to the SIE of the reference system.
If the reference system and test system resemble one another enough, we approximate that the same SIP will remove the same quantity of SIE in both.

Our reference systems for the O-SIPs, represented in Fig. \ref{fig:geometries}, are one-electron systems from our previous paper,\cite{Lonsdale2023} --- for which HF energies are exact for the given basis set. When talking about the datasets we bold and capitalize the name, e.g. \textbf{Linear 2}, \textbf{Triangle}, \textbf{Octahedron}, etc.

S-SIPs are made from a single hydrogen atom as the reference system.
We sequentially alter the contraction coefficient and Gaussian exponent to create a library --- we expound upon this shortly. 
We approximate that the effect the S-SIP has on the test system is similar to its effect to the H atom.
This is a very simple model and we do not expect this to perform well across all test systems, but will use any insights gained from this study for the constructions of future S-SIP generations.

An R script that allows using our SIP libraries are available for download on GitHub: \url{https://github.com/lgoerigk/SIPs}. Therein, the interested reader can also find all relevant information pertaining to the fits and spline interpolations, which are described next.

\begin{figure}
\begin{center}
\includegraphics[width=1\linewidth]{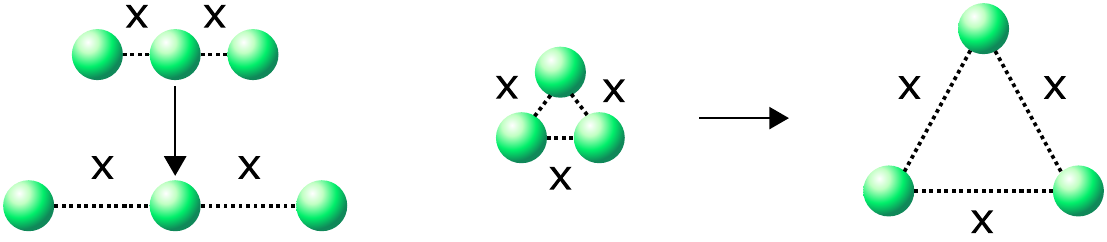}
\end{center}
\caption{Example of the relevant internuclear distance `x' in the \textbf{Linear 3} and \textbf{Triangle} systems such that they remain regular upon dissociation.}
\label{fig:symmetric_dissociation}
\end{figure}

\subsubsection{Optimized SIPs}\label{subsubsection:O-SIPs}
Four O-SIP libraries were constructed using the BLYP functional,\cite{BLYP_1_B88Ex,BLYP_2,BLYP_3} one from each of the following datasets: \textbf{Linear 2} $\rightarrow$ O-SIP-L2, \textbf{Linear 3} $\rightarrow$ O-SIP-L3, \textbf{Tetrahedron} $\rightarrow$ O-SIP-Tetra, and the \textbf{Octahedron} $\rightarrow$ O-SIP-Octa.
Each system has critical internuclear distances that we denote as `$x$'.
The definition of $x$ in \textbf{Linear 2} is simply the bond length between the two hydrogen nuclei.
For \textbf{Linear 3} $x$ is two bond lengths graphically represented in Fig. \ref{fig:symmetric_dissociation}.
This ensures polygons and polyhedra are regular --- $x$ is defined as the side length and this ensuring that the geometries conserve their symmetry: e.g. \textbf{$T_d$} for the \textbf{Tetrahedron} and \textbf{$O_h$} for the \textbf{Octahedron}.
Dissociation curves were made by uniformly increasing the distance `$x$' between each nucleus and its nearest neighbors.

All O-SIPs underwent two optimizations: the first was a free fit of both the Gaussian exponent and contraction coefficient (Eq. \ref{eq:SIP_equation}) to remove the SIE for each data point ($x$ distance), the second optimization was a partially constrained fit informed by the first optimization.
The optimization was a least-squares regression using the (SIE-free) HF energy as the reference energy.

During initial optimization of the SIP parameters the first data point, at $x=0.1$ {\angstrom} internuclear separation, was optimized first.
The resulting parameters that removed the SIE from this data point were then used as the starting parameters for the subsequent $x$ value.
This was repeated until the end of the dissociation curve.
Features of the initial `O-SIP-L2' and `O-SIP-Tetra' curves were used to construct the constraints of the second optimization.
Plots of these initial optimizations are found in SI Figs. S1 to S4.

In the second (and final) optimization of the O-SIPs, there were a series of constraints placed on the Gaussian exponent with respect to the internuclear separation, $x$.
This is summarized in SI Table. 1.
A more in-depth explanation of this procedure is given in SI Section 2.

Once these constraints were placed, the contraction coefficient was freely allowed to optimize to remove the SIE at the first data point, $x=0.1$ {\angstrom}.
Analogous to above, this coefficient was then used as the starting parameter for the next $x$ value until the end of the curve.
Plots for all O-SIP parameters are in the SI: O-SIP-L2 -- Fig. S7-8; O-SIP-L3 -- Fig. S9-10; O-SIP-Tetra -- Fig. S11-12; O-SIP-Octa -- Fig. S13-14.

Subsequently, all O-SIPs had both their contraction coefficients and the Gaussians exponents fitted to piece-wise linear splines.
These splines interpolated between each $x$ distance to yield smooth extraction of parameters.
Splines were chosen to be of degree 1 (linear) and the number of degrees of freedom was set equal to one less than the total number of data points.
We used R version 3.6.1 (2019-07-05) with the packages ``stats'' (version 3.6.1) and ``splines'' (version 3.6.1).

\subsubsection{Subtraction SIPs}

Generation of an S-SIP library is relatively simple: starting with an ECP of the form in Eq. \ref{eq:SIP_equation}, we alter the contraction coefficient, $c_{li}$, and the Gaussian exponent, $\zeta_{li}$, and then input them on an ECP into an all-electron calculation of the hydrogen atom.
The ECP changes the total energy and we record this as energy change due to the S-SIP:

\begin{equation}
\Delta E^{S-SIP}_{c,\zeta} = E^{Total}_{DFA+ECP} - E^{Total}_{DFA}\ .
\end{equation}

The reference energy is then the total energy of an unadulterated hydrogen atom at the same level of theory as the DFA$+$ECP calculation.
In a sequential fashion we alter the parameters $c_{li}$ and $\zeta_{li}$ and record the data into a library where it is fitted to a linear spline and given the title ``S-SIP''.

To choose the S-SIP that will correct the SIE we find the ECP-energy-change that is the additive inverse of the target system's SIE: $\Delta E^{S-SIP} = -SIE$.
In the ideal scenario this S-SIP `subtracts' the SIE from the total energy of the target system --- assuming we have a reliable measure of the SIE.
The assumption is we know how the S-SIP will interact with the target system --- the validity of which is tested in subsequent sections.

It was found convenient to create S-SIPs by holding the Gaussian exponent at a fixed value and only adjusting the contraction coefficient.
We named S-SIPs after the constant Gaussian exponent: S-SIP-0.001, S-SIP-0.1, S-SIP-1, S-SIP-2.5, S-SIP-20, S-SIP-100, and S-SIP-150.
The contraction coefficient ranged from $-$100 to 200 with a step width of $0.15$ --- with the exception of S-SIP-0.001 and S-SIP-150, which ranged from $-$175 to 174.8 with a step width of $0.3$.
All S-SIPs were constructed with BLYP; we trialed other DFAs as the base, but deemed this too far a digression away from the essence of this manuscript.

Lastly, a spline was fitted to the $\Delta E^{S-SIP}$ vs $c_{il}$ curve analogous to the O-SIPs in Section \ref{subsubsection:O-SIPs}.
This interpolation allowed coefficient values to be chosen when there was no explicit $\Delta E^{S-SIP}$ data point for a particular SIE.
Example fits are provided in SI Figs. S15-S19.

\subsection{Computational details}
\label{subsec:computational_details}

All calculations were conducted with Q-Chem V6.0.1\cite{QchemRef} and, for specific figures in the SI, ORCA 5.0.2.\cite{Neese2022}
The decontracted def2-QZVPP\cite{def2basen} basis set was used from the basis set exchange.\cite{Feller1996, Schuchardt2007, Pritchard2019} 
An unpruned grid composed of 99 radial spheres and 590 angular (Lebedev) grid points was chosen to eliminate grid-related dependencies.
All SIPs were constructed exclusively with the BLYP density functional.

Coupled cluster singles, doubles, and perturbative triples [CCSD(T)]\cite{Raghavachari1989} calculations with the decontracted def2-QZVPP basis set were carried out in Q-Chem V6.0.1 on H$_2$ and H$_3^{\ddagger}$ systems as per subsection \ref{subsubsec:sub_chapter_2d_H_abstraction}.

Excitation energies for the H atom were obtained with standard linear-response time-dependent DFT within the adiabatic approximation.\cite{Ullrich2011,Runge-Gross,TDDFTGrossKohn,CASIDA_TDDFT_1995,AdiabatApprox} 

DFAs used in this study include SVWN5,\cite{xalpha,VWN} BLYP, B97M-V,\cite{B97M-V} SCAN,\cite{regular_SCAN} TPSS,\cite{TPSS} B3LYP,\cite{B3LYP_1,b3lypb} cam-B3LYP,\cite{CAM-B3LYP} HFLYP (100 \% HF exchange and LYP correlation),\cite{BLYP_2} and $\omega$B97M-V.\cite{wB97M-V}

A quirk of Q-CHEM V6.0.1 is that ECPs cannot be placed on hydrogen atoms; nor on helium.
To sidestep this issue we replaced all H nuclei with Li nuclei.
We then placed a $-2$ point charge directly on the xyz-coordinates of each Li nucleus and adjusted the total charge to ensure the number of electrons was correct (one electron in most cases).
We also defined a custom basis set to ensure the correct H-atom decontracted def2-QZVPP was used on the lithium nuclei instead of the standard Li-basis set. An example QCHEM input can be found on our aforementioned GitHub.

Validation of our SIPs was conducted on four datasets that we detail below.

\subsection{Validation datasets}\label{subsubsec:validation_datasets}

\textbf{Geometry dataset}

The first of the validation sets is our geometry dataset (Fig. \ref{fig:geometries}).\cite{Lonsdale2023} 
Therein we chose 11 geometries and constructed them by placing hydrogen nuclei at each vertex, or in the case of the linear systems in a straight line,  represented in Fig. \ref{fig:geometries}.
Each model system possesses a single electron, hence Hartree-Fock is exact and serves as our reference energy.
Identical to the procedure explained in the O-SIP Subsection \ref{subsubsection:O-SIPs}, dissociation curves were created by increasing the internuclear separation, `$x$', between each H nucleus and its nearest neighbor(s) equidistantly.
Each of the 11 geometries made a dataset by ranging internuclear separation values from $x = 0.5$ to $x = 10\ \angstrom$ with a step width of $\Delta x = 0.1\ \angstrom$. 

\textbf{Excited states of hydrogen}

This is a slightly modified version of the higher orbital occupation dataset from our previous work.\cite{Lonsdale2023}
Herein we calculate the excitation energy of a hydrogen atom into one of the following orbitals: 2s, 2p, 3s, or 3d.
We calculate the error of the excitation with respect to Hartree-Fock as the SIE.
Applying SIPs in this case meant correcting the ground state energy to account for the erroneous excitation energy.

\textbf{Hydrogen transfer barrier height}

Lastly, the simple hydrogen transfer, $H_2 + H \rightarrow H + H_2$, is chosen as a many-electron validation.
We only take the barrier height of the transition state.
$H_2$ and $H_3^\ddagger$ geometries were taken from the BH76\cite{htbh38,nhtbh38,Goerigk2017}  set in the  GMTKN55\cite{Goerigk2017} database, whilst the reference energies were calculated with CCSD(T)/decontracted def2-QZVPP.

For the many-electron systems the SIE values were estimated (we use that term loosely) by taking the difference between the total energies of BLYP and HFLYP.
In short, if our SIPs worked perfectly they would correct the BLYP result to a HFLYP result.

\textbf{Hydrogenic mononuclear series}

This comes from our first study of the one-electron SIE.\cite{Lonsdale2020}
This is a dataset consisting of one-electron nuclei (cations) in series from element number 1 to 36.
Explicitly stated: H, He$^+$, Li$^{2+}$, Be$^{3+}$ $...$, Kr$^{35+}$.
We take the SIE to be the deviation from the HF reference energy.

\subsection{SIP construction}
\label{subsec:sip_construction}

A series of tests were conducted to guide the general form of the SIP. 
Relevant questions included what basis set to use, what grids were acceptable, how functional dependent they are, the number of ECP parameters, and finally which parameters to optimize or choose to hold constant.
Additional checks were done to ensure that the combined use of lithium nuclei, custom basis sets, point charges, and the overall charge to emulate hydrogen nuclei did not induce artifacts that could be interpreted as results. 
Preliminary tests revealed that the s-channel is sufficient for correcting the SIE and changing $n_{li}=1$ or $0$ only worsened results.
Various linear combinations of Gaussian functions were trialed successfully, but we invoked Occam's razor once it was clear that a single Gaussian was satisfactory.
Our final correction was minimalistic: consisting of only a single ECP in the s-channel that does not replace any electrons.

SI Section 4 shows that there is not a large grid dependence, nonetheless we opted for the equivalent of an unpruned SG-3 grid --- the largest default grid in Q-Chem 6.0.1.
We show in SI Section 5 that our chosen basis set of decontracted def2-QZVPP was adequate --- it was similar to the highly accurate hydrogenic basis sets\cite{Lehtola2020} (SI Fig. S27 and S28) and had better energetics (SI Figs.  S22-S25) than the decontracted aug-cc-pVQZ basis set we used in our first one-electron SIE study.\cite{Lonsdale2020}
A decontracted triple zeta basis set had undesirably large errors --- see SI Fig. S25 \& S26 --- and so we opted for a quadruple-$\zeta$ basis set.

To test if using lithium nuclei and adding point charges was the same as using hydrogen nuclei we created SI Fig. S29.
Error values of $\approx 2\times 10^{-7}$ $E_h$ demonstrate there is no substantial difference outside of the range of numerical noise. 
SI fig S30 shows that PBE\cite{PBE} did have larger issues for unknown reasons, and so we decided that PBE should not be included in this study.
Note that for Hartree-Fock there is no error, unlike for the DFAs BLYP, LDA, SCAN, and TPSS.
We have contacted the Q-Chem developers to rectify the problem that ECPs are not allowed for H and He, a problem other programs do not have. Thus, for the future we hope that the solution described earlier is just temporary for our development purposes.

\section{Self-interaction potentials performance}
\label{subsec:chapter2}

Here we evaluate each of the SIPs in terms of their performance on the aforementioned test sets.
We stress that our intention is to apply a SIP and move the energy by exactly the amount we desire --- this corresponds to the SIE in one-electron systems.
In such single electron cases any residual error that the SIP has not corrected can be considered either an under- or over-correction.
In the barrier height correction, success is not only determined by getting a value close to CCSD(T), but also by getting close to the chosen SIE-free reference, which is HFLYP and is explained in more detail in Subsection \ref{subsubsec:sub_chapter_2d_H_abstraction}. 

We define any remaining energy after the correction as the "residual":
\begin{equation}
\Delta E^{SIP}_{residual} = E^{total}_{DFA+SIP} - E^{total}_{reference},
\label{eq:delta_E_SIP_residual}
\end{equation}

where $E^{total}_{DFA+SIP}$ is the total energy of the SIP-corrected DFA calculation and $\Delta E^{total}_{reference}$ is the total energy of the reference Hartree-Fock calculation, or in the case of the many electron systems, the energy of HFLYP.
Ideally when applying a SIP there should be no residual, as accurately and reliably predicting the energy change from the SIP is key.
Upon achieving that goal, the only hurdle that remains is reliably predicting a quantity for the SIE in many electron systems --- also no easy task.

Note that the uncorrected SIE will be shown in many of the figures alongside the residual. 
For simplicity we will give the uncorrected values the same label by simplifying eq \ref{eq:delta_E_SIP_residual} into $\Delta E^{uncorrected}_{residual} =  E^{total}_{DFA} - E^{total}_{reference} = SIE$. Note we dropped the `SIP' as there is no SIP in an uncorrected calculation, even though figure labels will be $\Delta E^{SIP}_{residual}$.
In text the uncorrected residual we will simply be referred to as the SIE.

\subsection{SIP-corrected geometry dataset}
\label{subsubsec:sub_chapter_2a}

\begin{figure*}
\begin{center}
\includegraphics[width=1\linewidth]{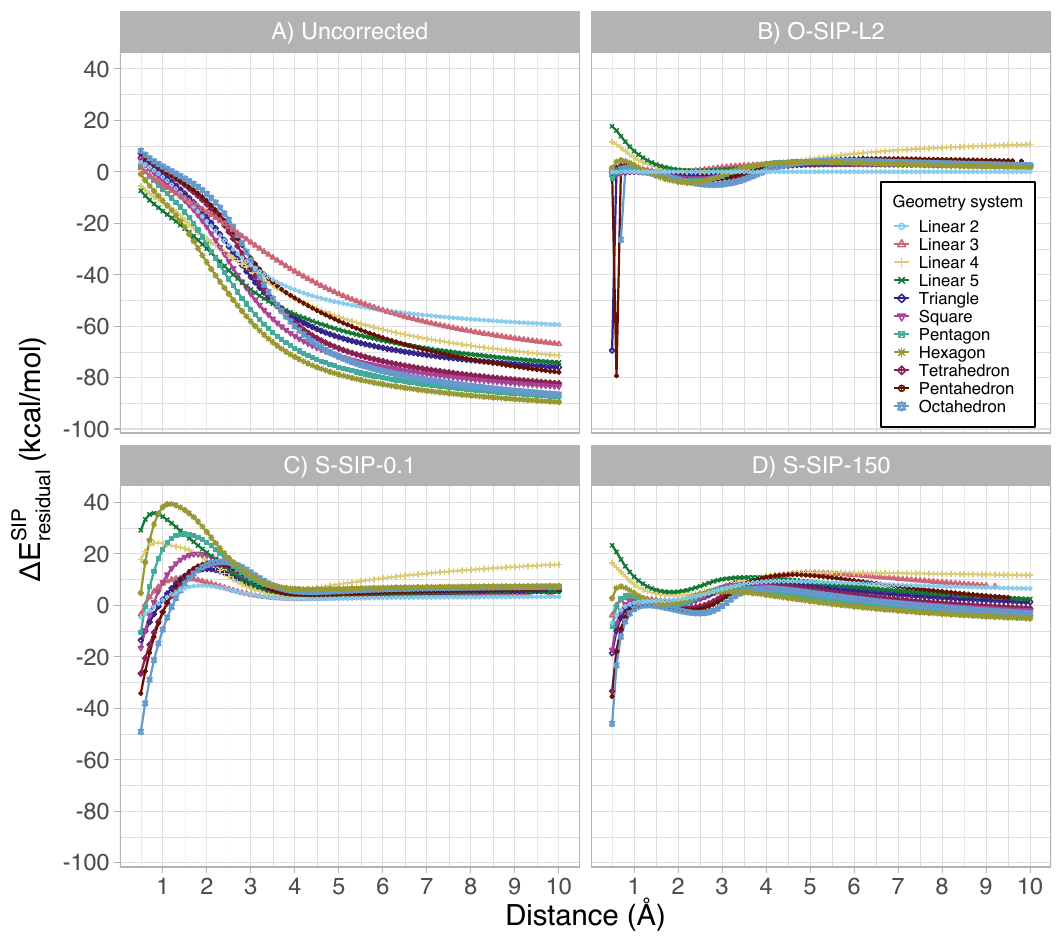}
\end{center}
\caption{Results for the geometry dataset: A) uncorrected BLYP, and BLYP corrected by B) O-SIP-L2, C) S-SIP-0.1, and D) S-SIP-150. The x axis refers to the distance `x' as defined in Fig. \ref{fig:symmetric_dissociation}. Basis set: decontracted def2-QZVPP.}
\label{fig:SIP_corrected_geometry_plots}
\end{figure*}

We begin evaluating performance of SIPs by using them to correct the geometry dataset.
Fig. \ref{fig:SIP_corrected_geometry_plots} shows the result of O-SIP-L2 correction, which is the best performing O-SIP.
O-SIP-Tetra and O-SIP-Octa corrected results can be found in the SI Figs.  S32 and S33 respectively --- O-SIP-L3 was created exclusively for section \ref{subsubsec:sub_chapter_2d_H_abstraction}.
Fig. \ref{fig:SIP_corrected_geometry_plots} also contains two of the better performing S-SIPs --- S-SIP-0.1 and S-SIP-150. 
Additional S-SIP results are in the SI Figs. S35-S41.
Only results pertaining to BLYP are shown in this section. 
Applying the BLYP-based SIPs to correct non-BLYP DFAs is still moderately successful; results can be found in SI Figs.  S31-S40.
Our plotted quantity, the residual $\Delta E^{SIP}_{residual}$, is the only source of error left in these datasets, if a residual has a value of 0 then it has completely removed the SIE from the corresponding data point.

Fig. \ref{fig:SIP_corrected_geometry_plots}A plots the uncorrected BLYP SIE to demonstrate the scope of the errors.
Panel B shows the result of applying O-SIP-L2 to each geometry dataset.
Verifying the optimization procedure worked as intended, the residual is essentially $0$ in the case of the \textbf{Linear 2} system.
On the whole these residuals from O-SIP-L2 are far superior compared to the uncorrected BLYP results.
Omitting outliers in the close range, corrected O-SIP-L2 results have errors ranging from $-5$ to $20$ kcal/mol, whereas the uncorrected BLYP extends from $-90$ to $10$ kcal/mol.
O-SIP-L2 corrected \textbf{Linear 4} and \textbf{Linear 5} possess the largest positive errors, other geometries do not have so high a range.

Outliers exist at very close internuclear distances ($<0.7\ {\angstrom}$) for the \textbf{Triangle}, \textbf{Pentahedron}, and \textbf{Octahedron}.
Such close proximity of multiple positively charged nuclei is a difficulty for the SIPs.
Crowded geometries are highly sensitive to the SIP parameters --- seemingly minor parameter variation becomes increasingly relevant as nuclei are forced together.
Additionally, convergence issues can arise when so many positive charges are nearby one another.
Generally both of these problems ease at greater internuclear separation.

Fig. \ref{fig:SIP_corrected_geometry_plots}B shows O-SIP-L2 tends to have an under-correction ($-\Delta E^{SIP}_{residual}$) between 1 and 5$\ \angstrom$, and over-corrections ($+\Delta E^{SIP}_{residual}$) seen below 1$\ \angstrom$ and beyond 5$\ \angstrom$.
In contrast, these two trends are generally reversed when O-SIP-Tetra is applied to geometry systems smaller than the \textbf{Tetrahedron} (Fig. S33). 
Similarly, these trends are fully reversed when applying O-SIP-Octa to the geometry set (Fig. S34) --- no system is smaller, except the \textbf{Hexagon} which is the same size.
We believe this has to do with the interplay of at least three factors: 1) SIPs are applied to all nuclei in the system, regardless of how many were used to make the SIP, e.g., O-SIP-L2 was designed on two nuclei, but the same SIP will be applied to three nuclei in the \textbf{Triangle} or \textbf{Linear 3};
2) there are different fractional charges, and therefore different delocalization errors, across the systems; 3) possible self-interaction linked Pauli repulsion --- the \textbf{Square} has less optimal spacing than the \textbf{Tetrahedron}.

\begin{figure}
\begin{center}
\includegraphics[width=1\linewidth]{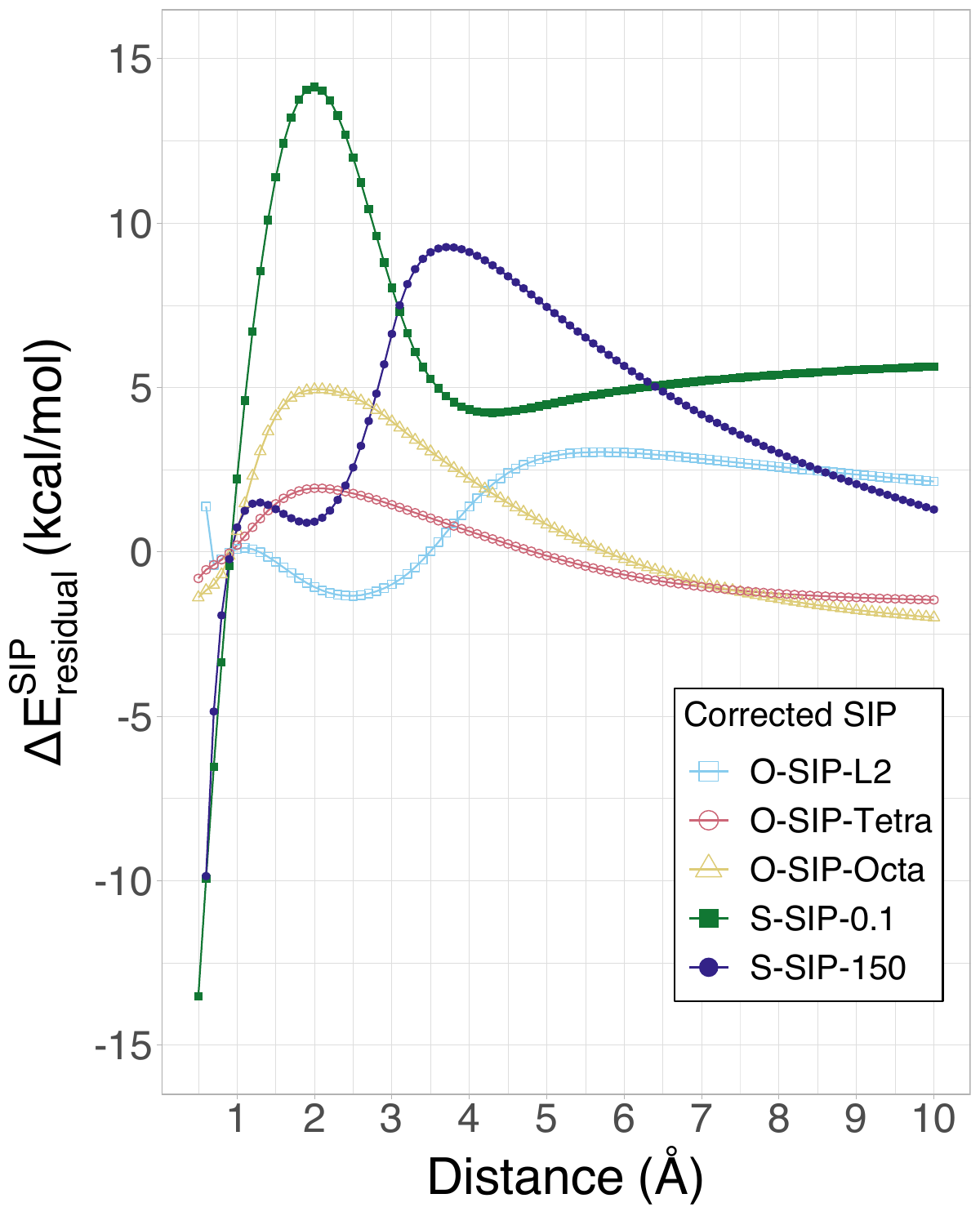}
\end{center}
\caption{Correcting the \textbf{Triangle} dataset (level of theory: BLYP/decontracted def2-QZVPP) with different optimized and subtraction SIPs. The x axis refers to the distance `x' as defined in Fig. \ref{fig:symmetric_dissociation}. Basis set: decontracted def2-QZVPP.}
\label{fig:triangle_corrected_plot}
\end{figure}

To more closely examine the observed trends of over- and under-correction, we created Fig. \ref{fig:triangle_corrected_plot}, which shows the residual errors for various SIPs when correcting the \textbf{Triangle}.
Therein we see O-SIP-L2 over-corrects the \textbf{Triangle} at long range ($>4\ \angstrom$).
Possibly this is because \textbf{Linear 2} has a fractional charge of 0.5 and is a dinuclear system, compared to the \textbf{Triangle} system which has an electron delocalized over three nuclei and a lower fractional charge per nucleus.
O-SIP-L2 is applying a dinuclear correction in full, which is probably excessive when three nuclei are present.
Compared to the mid range (1-5$\ \angstrom$) O-SIP-L2 under-corrects the \textbf{Triangle} likely due to \textbf{Linear 2} possessing a less severe positive SIE in this region.
It is clear to see O-SIP-Tetra reverses these effects, as does O-SIP-Octa.
Properties of the reference system (and therefore, the SIP) and the target system are in contention. 
The observed residual error, the magnitude of over- or under-correction, depends on the balance of properties between the reference system and the target.
To repeat verbosely, these properties unique to each system are the number of nuclei, their proximity (in angle and distance), and the fractional charge of each nucleus.

Without addressing each of these factors, we must accept a degree of residual error for this first generation of SIP libraries.
Unresolved contentions notwithstanding, the current residual errors are a substantial improvement over the uncorrected SIE --- comparing the SIE of panel A with the residuals from panel B (Fig. \ref{fig:SIP_corrected_geometry_plots}).

Within the SI we have included results from other DFAs. 
Figures S32 - S41 throughout SI Section 7 show a large reduction in the SIE for our other chosen DFAs: LDA, B97M-V, B3LYP, CAM-B3LYP, and $\omega$B97M-V.
Some of these DFAs behave rather differently in the dissociation of one-electron geometries,\cite{Lonsdale2023} but our SIP exhibits a robust functional independence for this system.
We believe this is because BLYP is one of the worst performing DFAs.
Therefore, choosing a BLYP-based SIP is beneficial as correcting other DFAs becomes a matter of interpolation and not extrapolation.
This assumes ``other DFAs'' have error ranges that are within the error ranges of BLYP.
Additionally, the above-explanation of the residual errors remains no matter the particular DFA in question; though the relative magnitudes of over- and -under-correction will (probably) change. 
We note functional dependence is much stronger at close internuclear distances ($\approx <1\ \angstrom$), and certain combinations of DFAs, systems, and internuclear separations are worse than others.

Subtraction SIPs S-SIP-0.1 and S-SIP-150 results are plotted in Fig. \ref{fig:SIP_corrected_geometry_plots}, panels C and D respectively.
These represent good performing S-SIPs with the remaining results being available in SI Figs. S35-S41.
Recall that these subtraction SIPs have a fixed exponent, in this case $\zeta = 0.1$ for S-SIP-0.1 (panel C) and $\zeta = 150$ for S-SIP-150 (panel D), of which we analyze S-SIP-0.1 first.
S-SIP-0.1 has a larger error range than O-SIP-L2, specifically at distances at and below $1.0$ \ \angstrom\ with residuals ranging from $-50$ to  $0$ kcal/mol. 
At distances beyond 2.5 $\ \angstrom$ the error range is typically much smaller with errors between 3 and 10 kcal/mol (with the exception of {\bf Linear 4}). 
Beyond $4.5\ \angstrom$ the residual tapers down to become a constant over-correction between $4$ and $8$ kcal/mol.
By comparison, S-SIP-150 (panel D) has a problem under-correcting distances below 1$\ \angstrom$, followed by an increase into a slight over-correction between $1-3\ \angstrom$ (varying by system).
In the short range, residuals range from $-$45 to 25 kcal/mol, but are worse than S-SIP-0.1 in the longer range with errors between $-8$ and 12 kcal/mol.

As internuclear distance increases, the residual S-SIP-150 slowly decreases --- likely due to the fact that 150 is relatively high exponent value representing a much more contracted Gaussian function than S-SIP-0.1.
S-SIP-0.001 in SI Fig. S35 shows extremely large over-corrected residuals in excess of $300$ kcal/mol at increasing internuclear separations --- possibly owing to the extremely diffuse Gaussian exponent value.

Fig. \ref{fig:triangle_corrected_plot} also shows S-SIP-0.1 and S-SIP-150.
In contrast to O-SIPs, S-SIPs approximate that the target system is similar in behavior to the H-atom.
With integer charge, a single nucleus, and no other nuclei, the H-atom is of poor equivalence to the \textbf{Triangle}.
O-SIP-L2 was based on the \textbf{Linear 2} ($H_2^+$) system that is made of two nuclei with fractional electron charge on each --- much closer to the \textbf{Triangle}.
Because the S-SIP reference system is so far removed from the target system, we do not know the relative proportions each of the mismatched features contributes to the overall residual.
We can, however, conclude that the fixed Gaussian exponent does matter. 
Hypothetically there are a multitude of parameters S-SIPs can be made from with no clear choice to which specific set to choose.
Fig. \ref{fig:triangle_corrected_plot} and SI Figs. S35-S41, while imperfect, clearly demonstrate that there is a subset of fixed Gaussian exponents that are better than others, subject to the distance.
Uncovering why will improve application of the next generation of S-SIPs.

Reflecting on the SIPs shown, corrections to 1D, 2D, and 3D one-electron model systems have been attained by both versions of SIP.
Within this test set are different fractional charges, regions of potential self-interaction based Pauli repulsion, number of nuclei, and functional- vs density- error based systems.
To that last point: \textbf{Linear 3}, \textbf{Linear 5}, and the \textbf{Pentahedron} were shown to suffer from density-driven SIEs in our last work.\cite{Lonsdale2023}
The rest of the systems shown in Fig. \ref{fig:geometries} suffer from functional-driven SIEs.\cite{Lonsdale2023}
S-SIP-0.1/150 and O-SIP-L2 are based on systems that have functional-driven SIEs, and yet they can adequately correct systems dominated by both error types.

So far we have found that $\Delta E^{SIE}_{residual}$ has a dependency on how similar the SIP is to the target system --- the more similar the better the correction.
With the data generated thus far we are poised to create another generation of SIP.
Despite possible improvements, the idea of SIPs have already shown decent promise by correcting the geometry test set as well as they have considering the circumstances. This clear from comparing the uncorrected results in Fig. \ref{fig:SIP_corrected_geometry_plots}A with the SIP-corrected ones in panels B-D. In the following sections we will continue to investigate the boundaries of the first generation SIPs.

\subsection{SIP-corrected hydrogen atom excitations}
\label{subsubsec:sub_chapter_2c_excited_states}

\begin{figure*}
\begin{center}
\includegraphics[width=1\linewidth]{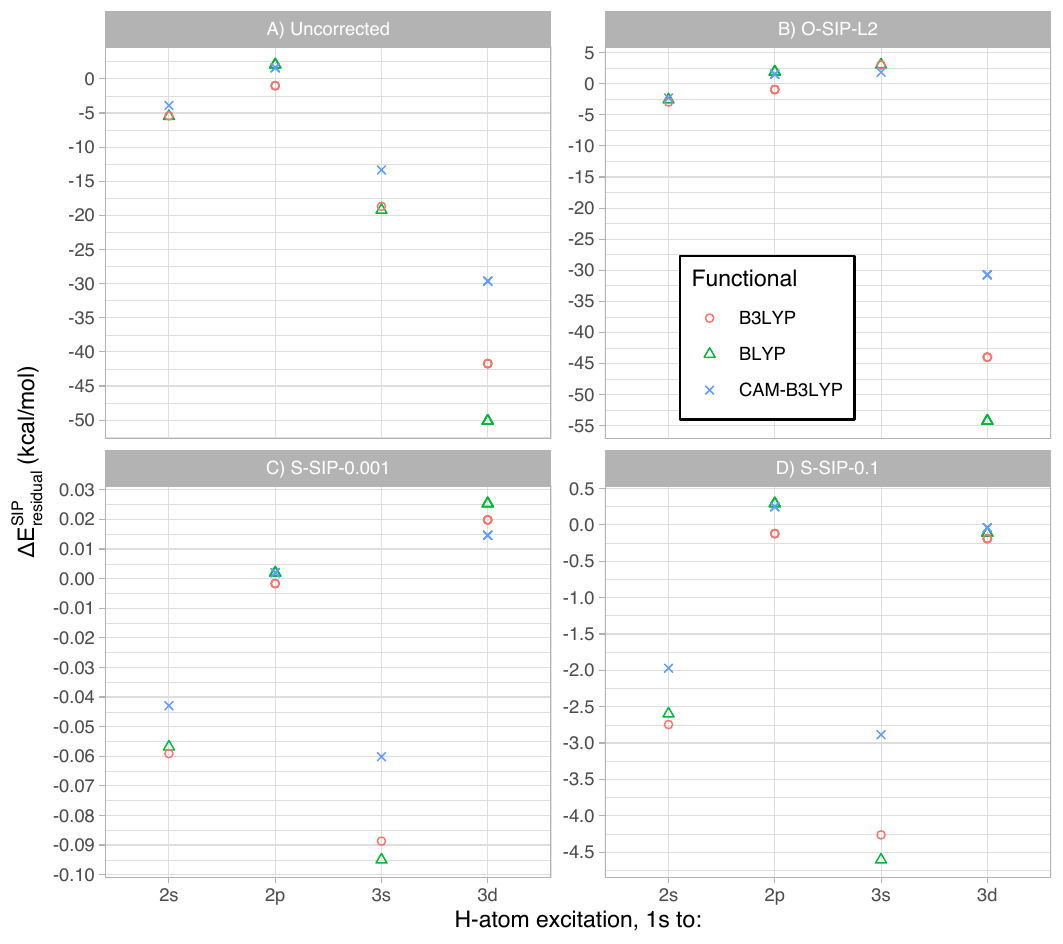}
\end{center}
\caption{SIEs and residual errors to excitation energies of the hydrogen atom for A) uncorrected case, and corrected by  B) O-SIP-L2, C) S-SIP-0.001, and D) S-SIP-0.1. Note, y-axes scale differently in each panel. Basis set: decontracted def2-QZVPP.}
\label{fig:excited_state_SIP_correction}
\end{figure*}

This test set involves correcting the excitations from the 1s ground to the 2s, 2p, 3s, and 3d excited states of the hydrogen atom. 
This is slightly modified from our previous foray\cite{Lonsdale2023} into converging an electron into higher lying orbitals using Q-CHEM's maximum overlap method,\cite{Gilbert2008} upon encountering technical issues when combining this technique with SIPs we converted to a simpler model. We determined the SIE to the excitation energy to chose the appropriate coefficient-exponent pair for the SIP that was then applied in the ground state and TD-DFT calculations.

Fig. \ref{fig:excited_state_SIP_correction}A shows the uncorrected results for BLYP, B3LYP, and CAM-B3LYP.
Errors range from SIE $\approx 2$ to $-50$ kcal/mol for BLYP, with the latter being the base functional all SIPs were designed on.
Corrected results are shown in the same figure: panel B shows O-SIP-L2 which is representative of the O-SIPs, panels C and D show S-SIP-0.001 and S-SIP-0.1 respectively, which are the best-performing SIPs for the this test set.
All other SIP-corrected results are available in the SI, Figs.  S42-S50.

O-SIP-L2 shows mild improvements for the 2s excitation (from SIE $\approx -5$ to $\Delta E_{residual}^{SIP} \approx -2.5$ kcal/mol) and decent improvement in the 3s excitation ($\approx -17.5$ to $\Delta E_{residual}^{SIP} +2.5$ kcal/mol).
There is essentially no change in 2p excitation error. 
There is a $-2$ to $-5$ kcal/mol increase in 3d excitation errors (e.g. SIE$^\text{BLYP} =-50$ kcal/mol to $\Delta E_{residual}^\text{BLYP/O-SIP-L2} \approx -55$ kcal/mol) --- the 3d excitations already having the largest errors to begin with.
Similar to our earlier argument but in reverse, current O-SIPs are designed using model systems with fractional electron charge, and therefore we do not expect particularly great performance on systems with integer electron charge.
Improvements in the s-type excitations is a positive result, the lack of performance in the 2p and 3d excitations may be attributed to the combination of relatively large exponent values of O-SIP-L2, combined with the fact that only the s-channel was used to make the O-SIPs.
Briefly jumping ahead of the results, S-SIPs also only utilized the s-channel in their construction but managed to correct the 2p and 3d excitations. It is possible the superior S-SIP performance is in spite of the mismatch of channels and not because of it.

Fig. \ref{fig:excited_state_SIP_correction}C and D show S-SIP-0.001 and S-SIP-0.1, respectively. 
Because of the drastic difference in residual errors, these two panels have different scales for the y-axes.
S-SIP-0.001 has a minuscule error range of $-0.10$ to $0.03$ kcal/mol corresponding to $-159$ to $47$ $\mu E_h$.
S-SIP-0.1 has more sizable residuals ranging from $-4.6$ to $0.25$ kcal/mol.
At first glance the excitations are into orbitals that are more diffuse than the 1s ground state, and so it follows that using a SIP with a smaller Gaussian exponent to correct it is physically reasonable.
This is also supported by the fact that both S-SIPs correct the 2p and 2d excitations better than the 2s and 3s excitations --- the former two involve more diffuse orbitals than the latter two.
However, S-SIP-0.1 is not anywhere near as good at correcting any excitation as S-SIP-0.001, suggesting that either $\zeta =0.1$ is too large an exponent for these tests, or that perhaps an exponent of $\zeta= 0.001$ is serendipitously accurate.
In the SI, Figs.  S44-S50 shows the other tested S-SIPs, also those with larger exponent values.
These figures reveal very similar performance to the O-SIPs, it appears they have exponent values that are too large to make much impact on the 2p and 3d excitations.

Partially inspired by our 2020 manuscript, Schwalbe, Trepte, and Lehtola published a continued analysis into ground and excited states of one-electron ions including a basis set study.\cite{Schwalbe2022}
Therein they showed that for the 2p and 3d excited states there is a larger basis set error --- though it is less pronounced for the hydrogen atom.
The current study shows that decontracted def2-QZVPP is a reasonable compromise between accuracy and time, residual errors as low as in S-SIP-0.001 presents in Fig. \ref{fig:excited_state_SIP_correction} ($<0.1$ kcal/mol), might be an indication that the the family of hydrogenic basis sets,\cite{Lehtola2020} or equivalent, might be required to eliminate basis set incompleteness errors in such a case. 
That being said, using such large basis sets to eliminate such a small error is not a feasible use of computational resources and we did not follow this philosophy here.

\subsection{SIP-corrected hydrogen transfer} \label{subsubsec:sub_chapter_2d_H_abstraction}

\begin{figure}
\begin{center}
\includegraphics[width=1\linewidth]{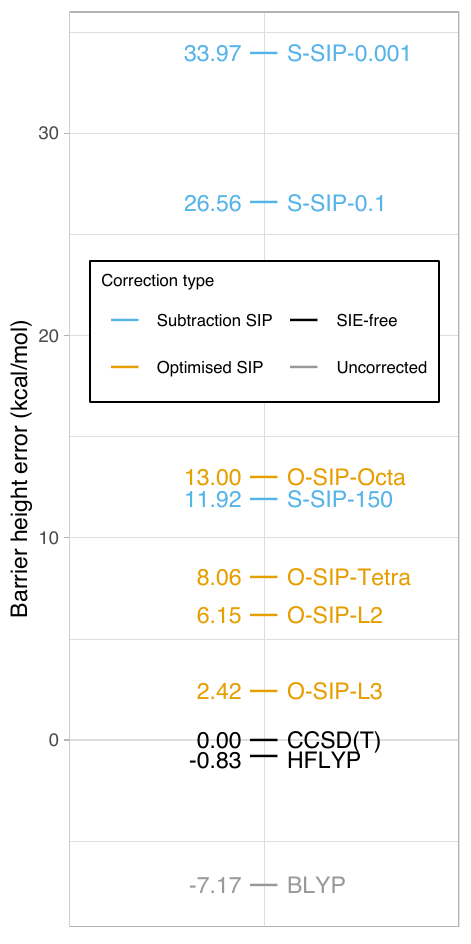}
\end{center}
\caption{Barrier height errors for the hydrogen transfer reaction $H_2 + H \rightarrow H + H_2$ relative to CCSD(T). Basis set: decontracted def2-QZVPP.}
\label{fig:optimised_and_subtraction_sip_barrier_height}
\end{figure}

For the many-electron test case the simplest hydrogen transfer reaction, $H_2 + H \rightarrow H + H_2$, is used as a test for correcting barrier heights.
An SIE-free wave function reference value was calculated with CCSD(T)/decontracted def2-QZVPP, yielding a barrier height of 10.10 kcal/mol. 
When choosing the right SIP, there is not yet a definitive way to quantify the SIE in many electron systems, therefore we used a rough proxy. 
We conducted two calculations, one with BLYP and another with HFLYP, and took the difference. 
Note that HFLYP is one-electron SIE-free.\cite{Lonsdale2020} 
This was done to correct each total energy of the three species involved in the reaction. 
The purpose of this test is not to correct the barrier per se, but to use a SIP to change the total energy of the system by a set amount. 
Once we can confirm we can accurately predict which SIP to use for which system, then the task of accurately estimating SIE values can be addressed. 
Firstly we analyze only our optimized SIPs.

Fig. \ref{fig:optimised_and_subtraction_sip_barrier_height} shows barrier height deviations with respect to CCSD(T) for uncorrected BLYP, and SIP-corrected methods. Note that HFLYP itself has a deviation of $-$0.83 kcal/mol from CCSD(T) due to differences in treating electron correlation effects. For this test specifically, O-SIP-L3 was created and included to document its performance in correcting the $H_3^{\ddagger}$ transition state. 

Our worst performer in the O-SIP category is O-SIP-Octa (deviation w.r.t. CCSD(T) = $+13.00$ kcal/mol), followed by the O-SIP-Tetra ($+8.06$ kcal/mol), O-SIP-L2 ($+6.15$ kcal/mol), and finally the best being O-SIP-L3 ($+2.42$ kcal/mol).
Uncorrected BLYP possesses a deviation of $-7.17$ kcal/mol w.r.t. CCSD(T).
All SIPs over-correct the barrier and only O-SIP-L2 and O-SIP-L3 provide better absolute deviations than uncorrected BLYP.

SI Section 7.4 has figures that show the residual errors with respect to HFLYP (the reference method we attempted to correct to).
It turns out that O-SIP-L3 is closest to HFLYP for the H$_2$ molecule and the H$_3^{\ddagger}$ transition state according to SI Figs.  S65 and S66.
In particular, correcting the transition state is energetically the most important as it is the greatest source of error.
In fact, those same SI figs show all O-SIPs, aside from O-SIP-L3, have error cancellation between the H$_2$ and H$_3^\ddagger$ systems.
This is good evidence that closer similarity between the target and the SIP-reference systems is beneficial to this first generation of SIPs.

Turning to S-SIPs in Fig.  \ref{fig:optimised_and_subtraction_sip_barrier_height} we consider their ability to correct the reaction barrier height.
Immediately we see that no tested S-SIP is better than uncorrected BLYP --- all barrier heights are too high.
Total energy of the H$_3^{\ddagger}$ transition state is severely over-corrected (11-32 kcal/mol) for all S-SIPs and is the primary source of error --- see SI Fig.  S69.
Total energies of the reactants reveal that the correction to the H atom has an insignificant error ($<0.02$ kcal/mol), while the H$_2$ energy is under-corrected (from $\approx -3$ to  $-2$ kcal/mol) according to SI Figs.  S65 and S66. 
In fact, H$_2$ energies are only worsened by applying a subtraction SIP --- the uncorrected energy is too low (by $-1.03$ kcal/mol), yet the subtraction SIP only further lowers the total energy for this system.
This yields error compensation between H$_2$ and H$_3^{\ddagger}$ which makes the S-SIP corrected barriers appear slightly better.
A consistent feature across this group is that the higher the fixed exponent, the better the total energies for H$_2$ and $H_3^{\ddagger}$ with S-SIP-150 being the best.
This follows given the higher exponents are more likely to be relevant for the orbitals that are occupied.

Applying S-SIPs in their current form does not translate well to real systems.
We believe this to be a result of not accounting for fractional charge and because they are based on singular H atoms rather than polynuclear systems.
While H$_2$ is a neutral molecule just like the H atom the S-SIPs were based on, it is clear blindly applying an S-SIP to this system does not yield the desired behavior.
One may expect that poor S-SIP performance on the H$_2$ molecule to be strange given both correction and the system possess integer electron charge.
Why the subtraction SIPs fail to work for H$_2$ may stem from the positive SIE that we quantified in previous work.\cite{Lonsdale2020,Lonsdale2023}
Typically the SIE across two nuclei is a negative value, but at closer internuclear distances (i.e. at the equilibrium bond length of H$_2$) there exists a positive-signed SIE.
The origin of this kind of positive SIE originates from the proximity of nuclei.
Because subtraction SIPs are based on a lone nucleus of integer charge their application to a strongly polynuclear system is wildly different than we approximate --- hence why S-SIPs fail when applied to H$_2$.

At a minimalistic glance, we can conclude that for this simple hydrogen transfer reaction we can apply a SIP and partially correct the total energy.
All of our optimized SIPs have shifted the total energies of the reactants/products and transition state in the correct direction.
More work is left in controlling the magnitude of the SIP correction.
Among other tests already mentioned, we have yet to account for the increased number of electrons that the hydrogen transfer introduces.
The first generation of S-SIPS do not seem to be competitive enough to handle this type of reaction and need further improvement.

\subsection{SIP-corrected hydrogenic mononuclear series}
\label{subsubsec:sub_chapter_2e_hydrogenic_mononuclear}

\begin{figure*}
\begin{center}
\includegraphics[width=1\linewidth]{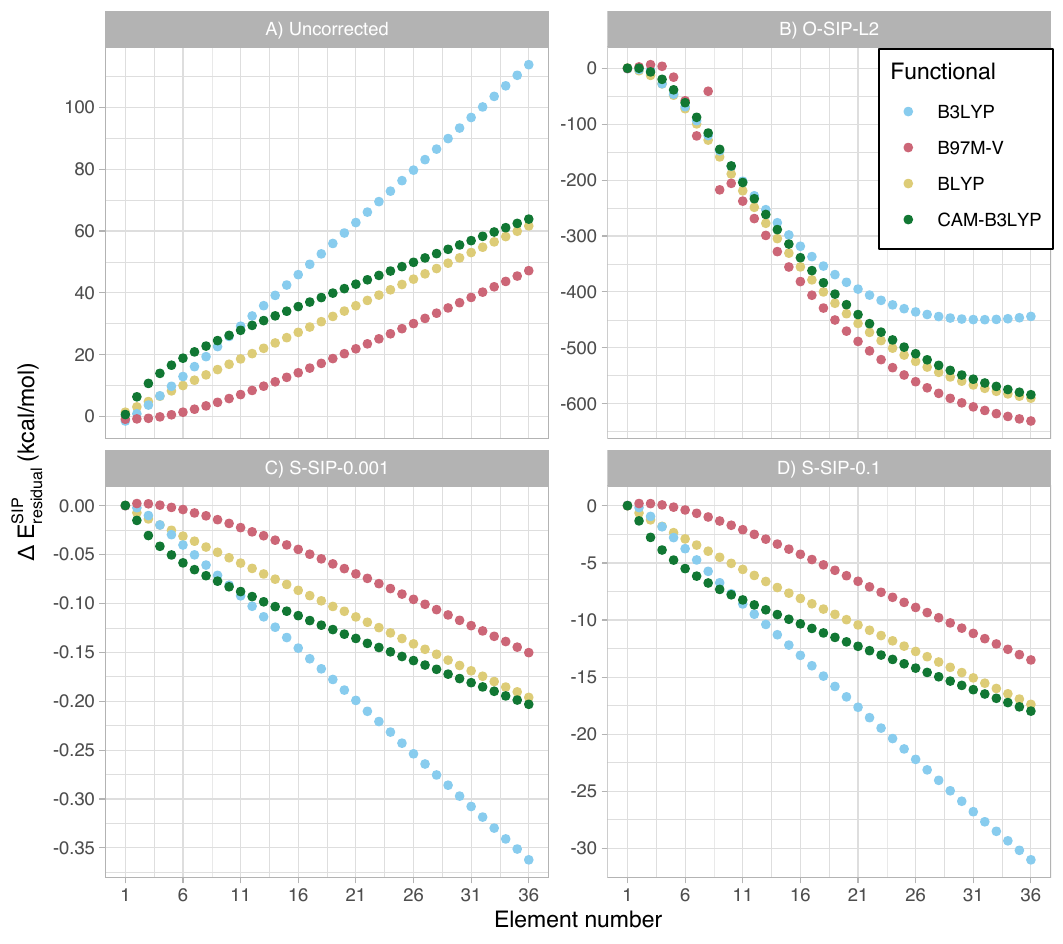}
\end{center}
\caption{Mononuclear hydrogenic dataset results: A) uncorrected BLYP, and BLYP corrected by B) O-SIP-L2, C) S-SIP-0.001, and D) S-SIP-0.1. Note, y-axes scale differently in each panel. Basis set: decontracted def2-QZVPP.}
\label{fig:hydrogenic_mononuclear_reference_and_optimised_sips}
\end{figure*}

The hydrogenic mononuclear series comes from our first publication on this topic\cite{Lonsdale2020, Lonsdale2023} consisting of 36 hydrogenic mononuclear systems in the series H, He$^+$, Li$^{2+}$, ... , Kr$^{35+}$.
Uncorrected SIE values for select DFAs are shown in Fig.  \ref{fig:hydrogenic_mononuclear_reference_and_optimised_sips} A, reminding the reader that the SIE increases in magnitude in an approximately linear fashion with increasing nuclear charge $Z$.\cite{Lonsdale2020}  The error range for uncorrected BLYP is 1.3 kcal/mol at H to 61.6 kcal/mol at Kr$^{35+}$

Coefficient-exponent pairs for the SIPs are chosen based on the SIE and the prior H-based SIP libraries --- i.e. element specific libraries were \textit{not} created for this test set.
We have also shown DFAs other than BLYP in this test set, even though all SIPs were created using BLYP, to test the functional dependence of SIPs.

Representative of our O-SIPs is O-SIP-L2 shown in Fig.  \ref{fig:hydrogenic_mononuclear_reference_and_optimised_sips}B.
Compared to the uncorrected values O-SIP-L2 over-corrects the SIE massively:  at Kr$^{35+}$ the BLYP residual is as large as $-600$ kcal/mol.
O-SIP-Tetra (Fig.  S52) and O-SIP-Octa (Fig.  S53) are not as egregious in their over-correction --- in fact, for three functionals (including BLYP) O-SIP-Octa has a better residual range than uncorrected.

Poor results of O-SIP-L2 may come from a mixture between O-SIPs being based on 1) fractional charge reference systems, 2) having multiple nuclei and 3) only possessing a nuclear charge of $Z=1$.
O-SIP-L2 is based on a system with a Mulliken spin density of $0.5$ associated with each H nucleus, while the O-SIP-Tetra and O-SIP-Octa are based on systems with atomic Mulliken spin densities of $0.25$ and $0.1\overline{6}$ respectively.
In one-electron cases, these population analyses correspond directly to the fractional charge on each nucleus.
If we recall when DFAs deviate from piece-wise linearity, a larger deviation occurs at half-integer fractional electron charge, than the deviation at $0.25$ (i.e. O-SIP-Tetra) or $0.1\overline{6}$ (i.e. O-SIP-Octa) fractional charge --- hence why the various O-SIPs had varying amounts of over-correction.
The greater the deviation from piece-wise linearity (with O-SIP-L2 being the largest) the larger the residual over-correction.

Our two best S-SIPs, S-SIP-0.001 and S-SIP-0.1, are shown in Fig.  \ref{fig:hydrogenic_mononuclear_reference_and_optimised_sips}C and D,  respectively. There is a substantial improvement over the uncorrected data with S-SIP-0.001 reducing the error by a factor of 50 ($\Delta E^{S-SIP} = -0.4$ to $0$ kcal/mol), and the S-SIP-0.1 reducing the error a factor of 3($\Delta E^{S-SIP} = -30$ to $0$ kcal/mol). 
The other SIPs with greater exponent values perform far worse to the point where the correction is orders of magnitude greater than the original SIE --- similar to the optimized SIPs.
At this point, it is expected S-SIPs are more suitable to the application of test systems with integer electron charge.

Akin to subsection \ref{subsubsec:sub_chapter_2c_excited_states}, excitation of the H-atom, the smaller exponents are much better at correcting the SIE than larger exponents.
This is contrary to our initial thought --- one may expect that a nucleus with higher Z would require larger Gaussian exponents (tighter orbitals) for their 1s electron.
While this may be true when designing basis sets, the SIE in this case is best corrected by a diffuse exponent value.
We are not yet sure as to why this is the case.

We would also like to note that $\omega$B97M-V, a range separated hybrid, has non-linear behavior in this dataset. 
SI Figs.  S54 and S62 show that this functional behaves much worse than the others.
A similar, poor performance is also present for SVWN5, though it is of the opposite sign.
Here we would like to remind the reader that all O-SIPs and S-SIPs predict a pair of parameters that remove the SIE for specifically BLYP.
Testing the transferability of the BLYP-based SIPs on $\omega$B97M-V and SVWN5, the resulting $\Delta E$ residual demonstrates an extremely strong functional and Z dependence.
In contrast, SI Figs.  S51-S53 and S55-S61 shows a much weaker functional dependence for B3LYP, B97M-V, and CAM-B3LYP.
Therefore, caution would advise that SIPs should be functional specific at this stage. That being said, S-SIP0.001 is currently the best candidate for this test set.

\section{Summary and outlook}
\label{sec:conclusions}

Two schemes of ``self-interaction potentials'' (SIPs), developed from effective core potentials (ECPs), have been proposed and subsequently tested on three one-electron test sets and a simple hydrogen transfer reaction.
The schemes are the optimized-SIPs (O-SIPs) and subtraction-SIPs (S-SIPs).
The former use polynuclear one-electron systems as the foundational ingredient, whereas the latter is based on the hydrogen atom.
Two test sets were taken from our previous prior two studies:\cite{Lonsdale2020, Lonsdale2023} dissociation of 1D, 2D, 3D structures containing hydrogen nuclei as per Fig.  \ref{fig:geometries}, and the hydrogenic nuclei from H to Kr$^{35+}$.  The investigation of excitation energies of the hydrogen atom into the 2s, 2p, 3s, and 3d shells was an extension of our previous work in Ref. \citenum{Lonsdale2023}. The barrier height of the $H_2 + H \rightarrow H+H_2$ hydrogen transfer reaction was taken from GMTKN55's BH76 database.

The O-SIPs perform much better on the geometry dataset and the hydrogen transfer reaction.
This is likely because O-SIPs are optimized on systems that have fractional electron occupation --- which is the same in aforementioned test sets.
The S-SIPs perform much better when correcting the excited states of hydrogen and the hydrogenic mononuclear test series.
This is likely because S-SIPs are based an integer electron charge model system --- just like those particular test sets.
If the SIP is based on a model system that is similar to the target system it will likely serve as a better correction.

It is very promising that our tests showed improvements for systems that suffer from either functional- or density-driven errors, pointing at the potential robustness of the idea.

Given ECPs are available in all major computational chemistry codes, their use as SIPs is broadly applicable without much, if any, further software development required for their use. Interested readers can find scripts that allow using  all our first-generation SIP libraries (and future generations) on our GitHub: \url{https://github.com/lgoerigk/SIPs}.
The end goal would be SIPs that can be applied easily and universally in combination with any functional favored by users.
This would allow functionals on the lower rungs of Jacob's ladder to be more reliable and further increase the accuracy of already accurate functionals.

Our study serves as successful proof of concept and our findings gave us insights into developing the next generation of SIPs.
For example, we believe incorporating the interplay between the fractional charge, the SIE, and the SIP to be important for consideration for a second-generation SIP.
An extension to many-electron systems --- reliably correcting the total energy of many electron systems is different to the one-electron model systems. 
Furthermore, current SIPs require a quantitative estimation of the SIE in order to be applied. 
Either such an estimation, or a suitable proxy, needs to be created and validated so that SIPs can be applied to practical examples. 
Ideally, the estimate will be fast and universal.
We have shown in various model systems the efficacy of SIPs when both of the above criteria are met and are confident a generalized correction is possible.

\section*{Supplementary Material}
Supporting information is available for download.
Therein one can find plots of initial and final SIP optimizations, grid dependence tests, basis set dependence tests, validation that our ECP workaround introduced no errors, and all additional results of applying the SIP corrections to each of the four test sets.

\section*{Data Availability Statement}

Data supporting the findings of this manuscript are available for download in the supporting information.
Raw data is and access to our SIPs can be found on our GitHub: \url{https://github.com/lgoerigk/SIPs}
Any additional raw data can be requested from one of the corresponding authors.

\section*{Acknowledgments}
Dale R. Lonsdale acknowledges an Australian Government Research Training Program Scholarship. The authors are thankful for the allocation of computing resources by the National Computational Infrastructure (NCI) National Facility within the National Computational Merit Allocation Scheme (Project No. fk5). 
This research was additionally supported by the Research Computing Services NCI Access scheme at The University of Melbourne.

\bibliographystyle{jcp}
\bibliography{ECP_paper}

\end{document}